# easyplater: the easy way to generate microplate designs deconvolved from multivariate clinical data

Avigail Taylor[1,*] and Micah P. Fletcher[1]

[1] Oxford-GSK Institute of Molecular and Computational Medicine, Nuffield Department of Medicine, Centre for Human Genetics, Roosevelt Drive, Oxford, OX3 7BN, United Kingdom.

* Corresponding author. E-mail: avigail.taylor@well.ox.ac.uk

## Abstract

**Summary:** Microplate-based ‘omic studies of large clinical cohorts can accelerate biomedical research, but experimental power and veracity are compromised when plate positional effects confound clinical variables of interest. Plate designs must therefore deconvolve positional and biological variation, but existing computational approaches still rely on manual intervention to ensure adherence to spatial constraints. Here, we present three complementary advances that reduce researcher effort. First, we propose a weighted, multivariate plate design score comprising a novel metric of spatial autocorrelation that rewards global separation of similar samples, and a penalty for local, variable-wise homogeneous regions. Second, we use a network-based approach to identify clinically similar samples, generate random layouts under the constraint that similar samples are allocated to distal wells, and select the top-scoring candidate as an initial layout. Lastly, we use an efficient sample-swapping search to improve this layout. We implemented this method in easyplater, an R package for generating 96-well plate designs that takes clinical data and variable weights as input and outputs the top-scoring layout in CSV, XLSX or HTML format. Overall, easyplater substantially reduces user intervention, outperforms existing methods, and facilitates robust plate design for large, plate-based ‘omic studies.

**Availability and Implementation:** easyplater is available from https://github.com/IMCM-OX/easyplater under the BSD-3-Clause license.
**Contact:** avigail.taylor@well.ox.ac.uk

## 1 Introduction

Microplate-based high-throughput ‘omic studies of large clinical cohorts are popular because they have potential to accelerate biomedical research, for example proteomic technologies such as Olink and SomaScan can enable fast biomarker discovery (Lundberg *et al.* 2011; Rohloff *et al.* 2014; Ferkingstad *et al.* 2021; Sun *et al.* 2023). However, microplates are prone to positional effects (Liang *et al.* 2013; Lilyanna *et al.* 2018; Mansoury *et al.* 2021) which, like other batch effects, can increase false positive and negative rates of biological signal detection when confounded with clinical variables of interest (Leek *et al.* 2010). Furthermore, adjusting for these confounding positional effects using post-hoc statistical methods can reduce power to detect signals (Goh, Wang and Wong 2017). Therefore, it is best to deconvolve positional and biological variation at the plate design stage of an experiment, thus avoiding confounding in the first place.

Methods for generating deconvolved layouts fall into three categories:

(1) The OlinkAnalyze R package function `olink_plate_randomizer` (Nevola *et al.* 2025) and the web-based PlateDesigner application (Suprun and Suárez-Fariñas 2019) both use random number generation as the basis for sample to well allocation.

(2) Well Plate Maker (WPM) and PLAID (Borges *et al.* 2021; Francisco Rodríguez, Carreras Puigvert and Spjuth 2023) are underpinned by constraint satisfaction algorithms whereby samples that are ‘similar’ to one another are randomly allocated to non-neighbouring wells according to locational rules set by the user. In WPM, the user chooses one clinical variable to positionally constrain. PLAID, in contrast, allows users to apply constraints to multiple variables, however the approach is geared towards well-balanced compound concentration experiments with no missing data, rather than to imbalanced clinical data with missingness, and its application in this setting is not shown. Both WPM and PLAID aim to output a plate design that meets user-specified location constraints, but no design is produced if they cannot be met, and this is more likely to happen as the number of clinical variables to constrain increases.

(3) In the OmixeR R package (Sinke, Cats and Heijmans 2021), multiple plate designs are generated and scored, and the best scoring design is outputted. Specifically, OmixeR performs randomized sample to well allocation multiple times (default 1000), and then outputs the layout with the smallest sum of absolute correlations between clinical variables and user-specified plate zones.

Ostensibly, OmixeR offers the most advanced approach to generating deconvolved layouts because: (i) it always produces an output, (ii) the output selected is the best from multiple attempts, and (iii) it works in a multivariate clinical setting. However, OmixeR’s metric for scoring layouts requires predefined plate-zones which are assessed for correlation with clinical variables. Consequently, correlation versus alternative partitions of the plate are ignored, and must be checked manually if considered important. In addition, clinical variables cannot be weighted by importance. Overall, then, OmixeR still requires too much user intervention to ensure designs adhere to spatial constraints.

Here, we aim to significantly reduce researcher-hours spent generating deconvolved plate designs. Our approach is built on three complementary advances. First, we observe that requiring a layout to deconvolve multiple variables from positional effects is equivalent to requiring that it is not spatially autocorrelated with respect to those variables, where spatial autocorrelation describes the extent to which a variable is correlated with itself through space (Moraga 2024). Motivated by this observation, we propose a weighted, multivariate plate design score comprising a novel metric of spatial autocorrelation that rewards global separation of similar samples, together with a penalty for local, variable-wise homogeneous regions. Second, we use a network-based approach to identify clinically similar samples, generate constrained random layouts that preferentially separate similar samples across the plate, and select the top-scoring candidate as the initial layout. Finally, we use an efficient sample-swapping search to refine this layout according to the plate design score. We have implemented our method in easyplater, an R package that substantially reduces user intervention in plate design and consistently achieves greater deconvolution of clinical and positional variation than existing tools. As large, well-phenotyped cohorts become available for high-throughput ‘omic studies and numbers of plates and clinical variables grow, easyplater represents an important advance in automated plate design.

# 2 Methods

## 2.1 Plate design score

Our plate design score, $PDS$, must capture the requirement that, globally, samples in neighbouring wells do not have similar clinical variables, i.e., that designs do not have positive spatial autocorrelation with respect to clinical variables. In addition, $PDS$ should incorporate a local condition that variable-wise homogenous rows, columns and $3*3$ well-patches (herein termed ‘patches’) should be avoided. This is necessary because one layout can have less global spatial autocorrelation than another, whilst also having more local homogeneity than it (Supplementary Fig. S1), and $PDS$ should enable the latter layout to score more favourably than the former, if required. Therefore, we construct $PDS$ as the weighted sum of two sub-scores, $PDS_{global}$ and $PDS_{local}$, which respectively capture these global and local requirements.

Precise definitions follow, but for the moment we note that our construction of $PDS$ is predicated on two constraints. First, we assume that the same set of clinical variables is recorded for each sample to be plated; missing data can be tolerated, but less is better. Second, that the variables are categorical. Whilst it is possible to use numerical variables as-is, keeping similar samples apart is a stronger requirement than keeping identical samples apart; we therefore suggest binning numerical variables into ranges to treat them as categorical.

We now propose a construction for $PDS$ which incorporates multivariate data, and which allows users to weight clinical variables by importance.

### 2.1.1 $PDS_{global}$

The algorithm for calculating $PDS_{global}$ is underpinned by a domain-specific framework, which we set out here. We start by defining well-neighbourhood in a microplate, and describe positive to negative spatial autocorrelation within this context: For a given well, we define its neighbours as all the wells with which it shares a row or column (Fig. 1A), reflecting the tendency of whole rows and columns of wells to be collectively affected by the same positional effects on a microplate (Liang *et al.* 2013; Lilyanna *et al.* 2018; Mansoury *et al.* 2021). Then, if we have an $N$-well plate, two types of samples - say light and dark, and a requirement to fill $N/2$ wells with dark samples, and $N/2$ wells with light samples, we can describe the following patterns of spatial autocorrelation: clustering light and dark samples into two homogenous regions yields extreme positive spatial autocorrelation (PSA) wherein a pattern of similar neighbours dominates throughout (Fig. 1B); a chessboard layout exhibits extreme negative spatial autocorrelation (NSA) in which similar samples are maximally dispersed such that a pattern of dissimilar neighbours dominates throughout (Fig. 1C); all other layouts sit on a spectrum of PSA to NSA, exhibiting varying degrees of either similar or dissimilar neighbours dominating (Fig. 1D) (Radil 2011; Moraga 2024).

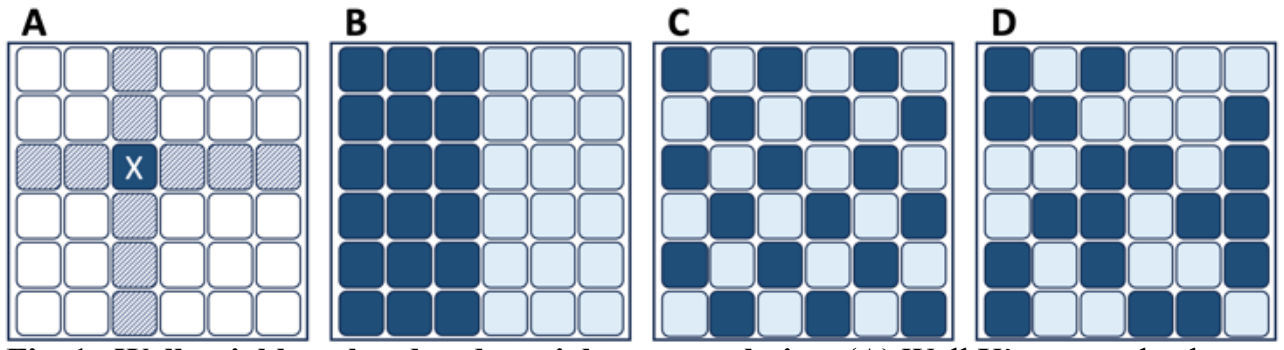

**Fig. 1. Well-neighbourhood and spatial autocorrelation. (A)** Well X’s row and column neighbours are indicated with hatching. **(B)** Extreme positive spatial autocorrelation. **(C)** Extreme negative spatial autocorrelation. **(D)** A layout which tends towards NSA: the pattern of dissimilar neighbours is more prevalent than the pattern of similar neighbours, but not dominant throughout. Figure adapted from (Radil 2011).

Now we define the range of values that $PDS_{global}$ takes. In particular, we require that for any given set of clinical, categorical variables, considering the values they take, the number of and size of the groups of samples taking each of those values, and the size of the microplate: plate designs approaching extreme PSA with respect to those variables score $PDS_{global} \rightarrow 0$; designs approaching extreme NSA score $PDS_{global} \rightarrow 1$; and all other designs score $0 < PDS_{global} < 1$ depending on their degree of spatial autocorrelation from PSA to NSA, with a score of 0.5 indicating that neither PSA nor NSA dominates the design. Putting all this together, we use the following algorithm to calculate $PDS_{global}$ for a layout where $N$ samples are assigned to wells in an $N$-well plate:

For each $v \in V$, where $V$ is the set of variables recorded for the $N$ samples, calculate $PDS_{global_v}$, the sub-score of $PDS_{global}$ measuring the degree of spatial autocorrelation of variable $v$ across the plate, as follows:

For each $x \in X_v$, where $X_v$ is the set of values that $v$ can take, calculate $PDS_{global_{v_x}}$, the sub-score of $PDS_{global_v}$ measuring the spatial configuration of value $x$ across the plate:

Consider the set of all pairs of samples where both samples have value $x$ for variable $v$, then:

- Count $t_{v_x}$, the number of such pairs of samples assigned to row and column exclusive wells.
- Calculate $t_{v_x_{min}}$ and $t_{v_x_{max}}$, the minimum and maximum number of such pairs of samples, respectively, that could feasibly be assigned to row and column exclusive wells, given the number of samples with this value and the dimensions of the plate.
- Then, $PDS_{global_{v_x}} = (t_{v_x} - t_{v_x_{min}})/(t_{v_x_{max}} - t_{v_x_{min}})$.

(See Supplementary Box S1 for further information.)

Once we have calculated $PDS_{global_{v_x}}$ for each $x \in X_v$, we have:

$$PDS_{global_v} = median(\{PDS_{global_{v_x}} : x \ in \ X_v\}) ,$$

and, finally:

$$PDS_{global} = \sum_{v \in V} PDS_{global_v} \cdot w_v / \sum_{v \in V} w_v ,$$

where $w_v$ is the user-assigned weight for $v$.

### 2.1.2 $PDS_{local}$

The condition that variable-wise homogeneous rows, columns and patches should be avoided in a layout is captured by the sub-score $PDS_{local}$. This is defined as an inverse penalty-based score where designs score maximally if they have a non-homogeneous distribution of values for every variable in every row, column and patch, but deviations from this are penalized. In particular:

$$PDS_{local} = \frac{\left(PDS_{row} + PDS_{col} + w_{pat} \cdot (PDS_{pat})\right)}{\left(PDS_{row_{max}} + PDS_{col_{max}} + w_{pat} \cdot (PDS_{pat_{max}})\right)},$$

where:

$$PDS_{row} = \sum_{v \in V} w_v \left(T_r - \left|\left\{\begin{matrix} r : r \in rows \\ \wedge\ |\{value(s,v) : s \in samples(r)\}| = 1 \end{matrix}\right\}\right|\right),$$

$$PDS_{col} = \sum_{v \in V} w_v \left(T_c - \left|\left\{\begin{matrix} c : c \in columns \\ \wedge\ |\{value(s,v) : s \in samples(c)\}| = 1 \end{matrix}\right\}\right|\right),$$

$$PDS_{pat} = \sum_{v \in V} w_v \left(T_p - \left|\left\{\begin{matrix} p : p \in patches \\ \wedge\ |\{value(s,v) : s \in samples(p)\}| = 1 \end{matrix}\right\}\right|\right),$$

$$PDS_{row_{max}} = \sum_{v \in V} w_v T_r,$$

$$PDS_{col_{max}} = \sum_{v \in V} w_v T_c,$$

$$PDS_{pat_{max}} = \sum_{v \in V} w_v T_p,$$

and:

- $rows$, $columns$, and $patches$, are the sets of row, column and patch indices in the microplate under design (Supplementary Fig. S2);
- $T_r$, $T_c$, $T_p$ are the total number of rows, columns and patches in the microplate, (so $T_r = |rows|$, $T_c = |columns|$, $T_p = |patches|$);
- $value(s,v)$ returns the value of variable $v$ for sample $s$;
- $samples(y)$ returns the set of samples allocated to wells in the row, column or patch indexed by $y$;
- $w_{pat}$ is a down-weighting for $PDS_{pat}$, required because $|patches| \approx 3(|rows| + |columns|)$. ($w_{pat} = \frac{1}{6}$ by default, Supplementary Fig. S2);
- $w_v$ is the user-assigned weight for $v$ (as before).

### 2.1.3 *PDS*

We can now define:

$$PDS = PDS_{global} + w_{local} \cdot PDS_{local},$$

where $w_{local}$ is the weight that lets us adjust the relative importance of $PDS_{local}$ to $PDS_{global}$. ($w_{local}$ =1 by default).

## 2.2 Generating a deconvolved plate design

With *PDS* in hand, we can now write down our algorithm for generating a deconvolved plate design for a set of samples with multiple associated clinical variables. Overall, our algorithm has three steps:

(1) Find communities of variable-wise similar samples.

(2) For $i = 1..n$ iterations (e.g., $n = 10$), randomly allocate samples to wells under the constraint that similar samples should be allocated to distal wells, and calculate *PDS* for the iteration's design.

(3) Taking the best scoring plate design from (2), search for randomized sets of sample switches yielding a plate design with improved *PDS*.

Here are the details:

**Step 1: Find communities of variable-wise similar samples**

For each pair of samples, *a* and *b*, we define their similarity, $s_{ab}$, as the weighted overlap of their clinical variables:

$$s_{ab} = \sum_{v \in V} w_v \cdot equal(a,b,v) / \sum_{v \in V} w_v,$$

and

$$equal(a,b,v) = \begin{cases} 1, & value(a,v) = value(b,v) \\ 0, & otherwise \end{cases},$$

where notation is as above.

Next, we construct a network of samples in which samples are represented as nodes and edges connect pairs of samples, *a* and *b*, where $s_{ab}$ exceeds some pre-defined threshold. Finally, we identify communities of similar samples using the Edge-Betweenness algorithm (Girvan and Newman 2002). Starting with the whole network as one community, this method repeatedly divides the network into smaller communities by iteratively removing edges with highest "betweenness", some measure which preferentially scores inter-community edges over intra-community edges. Here, where it is desirable to capture as much similarity as possible amongst our samples, this divisive approach to community detection is preferrable to agglomerative methods which perform well at identifying strongly linked interior nodes of communities, but may exclude outer nodes, even when those nodes do belong to a community (Girvan and Newman 2002; Newman and Girvan 2004; Smith *et al.* 2020).

**Step 2: Randomly allocate similar samples to distal wells**

In this step, we first calculate all pairwise well distances on the microplate for which we are designing a layout (Supplementary Fig. S3). Next, we iterate through the following process $i = 1..n$ times ($n = 10$ by default.):

In random order, for each non-singleton sample-community, *c*, found in Step 1:

- Identify all well-sets wherein all pairwise well distances exceed some threshold (Supplementary Box S2), and where there are the same or fewer wells than samples in *c* not yet assigned to a well.
- Randomly choose one of these well-sets, and allocate samples from *c* randomly to these wells.
- Repeat until all samples in *c* have a well allocation.

To finish, randomly allocate singleton samples to the remaining wells on the microplate, then calculate *PDS* for this design.

**Step 3: Search for an improved plate design using sample switches**

Now, we take the best scoring layout from Step 2, $D_0$, and generate sets of mutually exclusive sample switches which can be performed in one step, simultaneously, to make a new layout (Supplementary Box S3). We then employ this simultaneous switching step in a limited search through plate design space for a layout with improved *PDS*:

- With $D_0$ as the root node of our search tree, and corresponding score $PDS_{D_0}$, we use simultaneous switching up to $k$ times to generate up to $j$ new layouts (where $j \leq k$), $D_{0_1} \ldots D_{0_j}$, for which $PDS > PDS_{D_0}$.
- We then recursively repeat this search on layouts $D_{0_1} \ldots D_{0_j}$ up to a predefined depth of search, $M$, stopping at an earlier depth if no improved designs are found there.
- Finally, we output the best plate design found in our search space, that is, the layout which scored the highest *PDS*.

## 3 Results

To assess the performance of easyplater we used it to generate 50 plate designs for five synthetic datasets each containing 96 samples. In the first four datasets, samples had one of two values, "*v1*" or "*v2*", for one variable, *V*, with the proportion of samples having one or other of the values varying across datasets such that the ratio of *v1*:*v2* was 50:50, 60:40, 70:30, and 80:20. We also synthesized a fifth dataset with 96 samples and four variables, *A*, *B*, *C*, and *D* (Supplementary Fig. S4), and used easyplater to generate 50 plate designs with A to D equally weighted, and 50 designs with weights 0.1, 0.65, 0.15, and 0.1 assigned to the variables in order. For comparison, we repeated this process using a random number generator to allocate samples to wells, and OmixeR run in *row* mode (OmixeR(row)) and *column* mode (OmixeR(col)), i.e., minimizing the sum of absolute correlations between variables and rows, and variables and columns, respectively (Sinke, Cats and Heijmans 2021). (Supplementary Table S1 shows mean runtimes on a Bioinformatics-capable laptop.)

Looking at the distributions of *PDS*s obtained for generated designs, easyplater outperformed the other methods in every test, even when sub-scores $PDS_{global}$ and $PDS_{local}$ were considered separately (Fig. 2 and Supplementary Fig. S5). Since easyplater is designed to search for plate layouts with improved *PDS*, this result was expected; step 2 of the algorithm making the primary contribution to this improvement, with step 3 sometimes yielding a further, small increment in *PDS* (Supplementary Fig. S6).

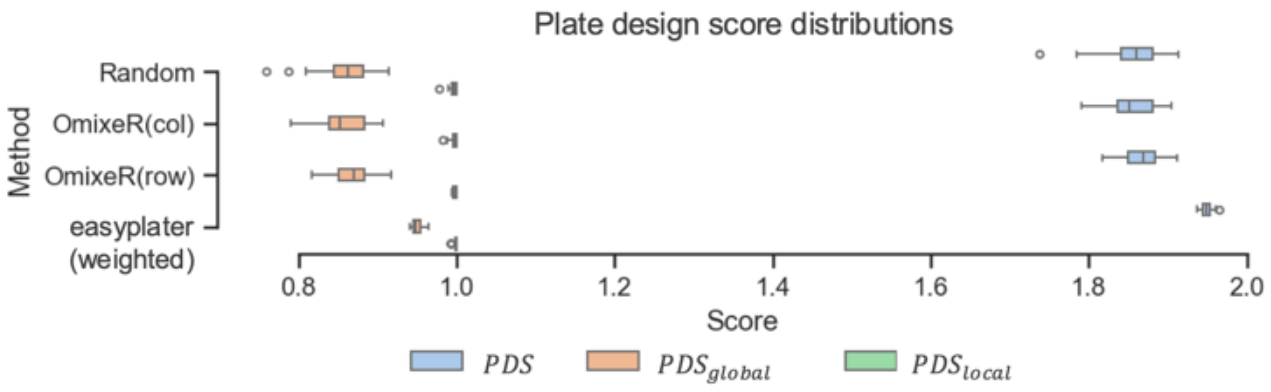

**Fig. 2. Distributions of plate design scores obtained for plate designs generated using random number generation, OmixeR, and easyplater.** Each method was used to make 50 plate designs for a synthetic, 96-sample dataset wherein each sample had four variables, *A*, *B*, *C*, and *D* (Supplementary Fig. S4). For easyplater, weights 0.1, 0.65, 0.15, and 0.1 were assigned to the variables in order (option not available in remaining methods). Random = random number generator, OmixeR(col) = OmixeR *column* mode, OmixeR(row) = OmixeR *row* mode; blue, orange, and green boxplots show distributions of $PDS$, $PDS_{global}$, and $PDS_{local}$, respectively.

We wanted to understand why OmixeR-generated layouts, in contrast to easyplater-generated layouts, yielded *PDS*-distributions overlapping those obtained for designs generated using random number sample to well allocation (Fig. 2 and Supplementary Fig. S5). To explore this, we looked at the relationship between the measures of spatial autocorrelation used by the two algorithms. Since easyplater tries to maximize *PDS,* whilst OmixeR seeks to minimize the sum of absolute correlations between variables and plate zones, we would expect a strong negative correlation between these measures if they were proxies of one another. However, although significant, only a weak negative correlation was observed (*PDS* vs. OmixeR(row): (Spearman's $\rho = -0.21$, $p < 0.001$); *PDS* vs. OmixeR(col): (Spearman's $\rho = -0.15$, $p < 0.001$); see Supplementary Fig. S8A & B), suggesting that the measures capture differing spatial properties of plate designs. Further, we found that OmixeR(row) and OmixeR(col) scores did not correlate with one another (Spearman's $\rho = 0.023$, $p = 0.476$. Supplementary Fig. S8C), but that -in contrast- *PDS* did correlate with their joint distribution ($p$ = 3.8e-05, Chi-square GOF test; Supplementary Fig. S8D). Overall, these results suggest that easyplater outperformed OmixeR in our tests because *PDS* is better at capturing NSA in plate design than correlation of variables with a single, prespecified gradient across a layout.

## 4 Using easyplater

The easyplater package ingests a tabular sample manifest and outputs both a 96-row sample manifest and a corresponding 96-well plate layout in CSV, XLSX or HTML format. Users can optionally specify wells to leave empty, for example for plate controls, and *PDS* calculations are automatically adjusted for the effective plate size. easyplater requires R version ≥ 3.5 and can be installed from https://github.com/IMCM-OX/easyplater, where a vignette demonstrating its use is also available.

## Acknowledgments

We would like to thank Nikoleta Vavouraki, Ayan Ianniello, and Georgia Brennan for their helpful discussions about the plate design problem, algorithm development and effective communication of the easyplater solution.

## Author contributions

AT (Conceptualization [lead], Methodology [lead], Software [lead], Writing [lead], Formal analysis [lead], Supervision [lead], Project administration [lead]), MF (Software [supporting], Visualization [lead]).

## Funding

This work has been supported by the Oxford-GSK Institute of Molecular and Computational Medicine.

*Conflict of Interest:* none declared.

# easyplater: the easy way to generate microplate designs deconvolved from multivariate clinical data

## Supplementary Information

Avigail Taylor[1,*] and Micah P. Fletcher[1]

[1]Institute of Molecular and Computational Medicine, Nuffield Department of Medicine, Centre for Human Genetics, Roosevelt Drive, Oxford, OX3 7BN, United Kingdom.

* Corresponding author. E-mail: avigail.taylor@well.ox.ac.uk

## Table of Contents

## Supplementary Figures

**Supplementary Figure 1. Global spatial autocorrelation versus local homogeneity in plate designs**

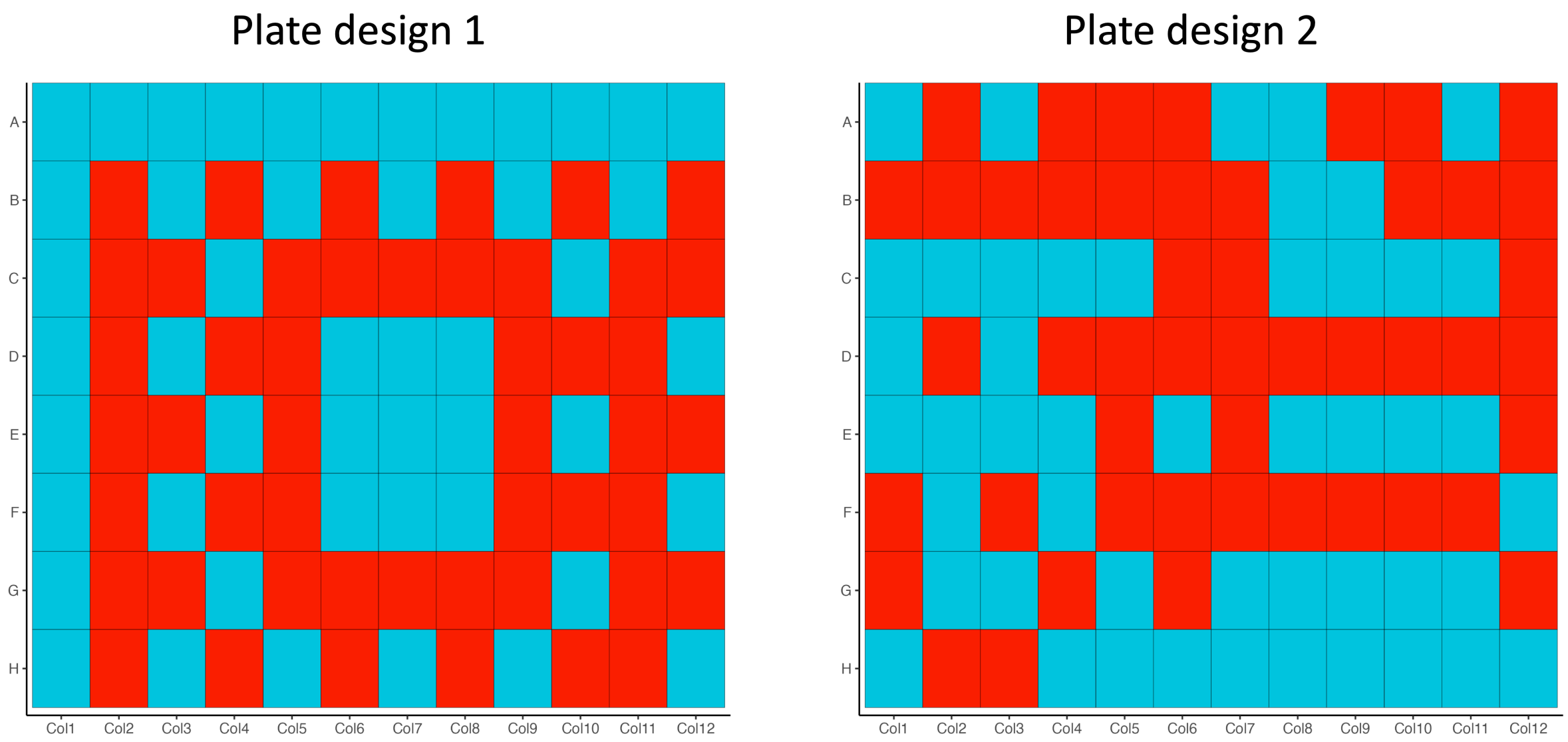

**Supplementary Fig. 1.** Here we show two plate designs for distributing 48 red and 48 blue samples across a 96-well microplate. Using the global measure of spatial autocorrelation described in section 2.1.1 of the main paper, $PDS_{global}$, Plate design 1 obtains a better, lower score ($PDS_{global} = 0.69$) than Plate design 2 ($PDS_{global} = 0.66$). Nonetheless, Plate design 1 has three regions of local homogeneity (row A: all blue, column 1: all blue, 3 x 3 well-patch (D6, D7, D8, E6, E7, E8, F6, F7, and F8): all blue), whereas Plate design 2 has no such regions.

**Supplementary Figure 2. Mathematical row, column, and patch indices on a 96-well plate**

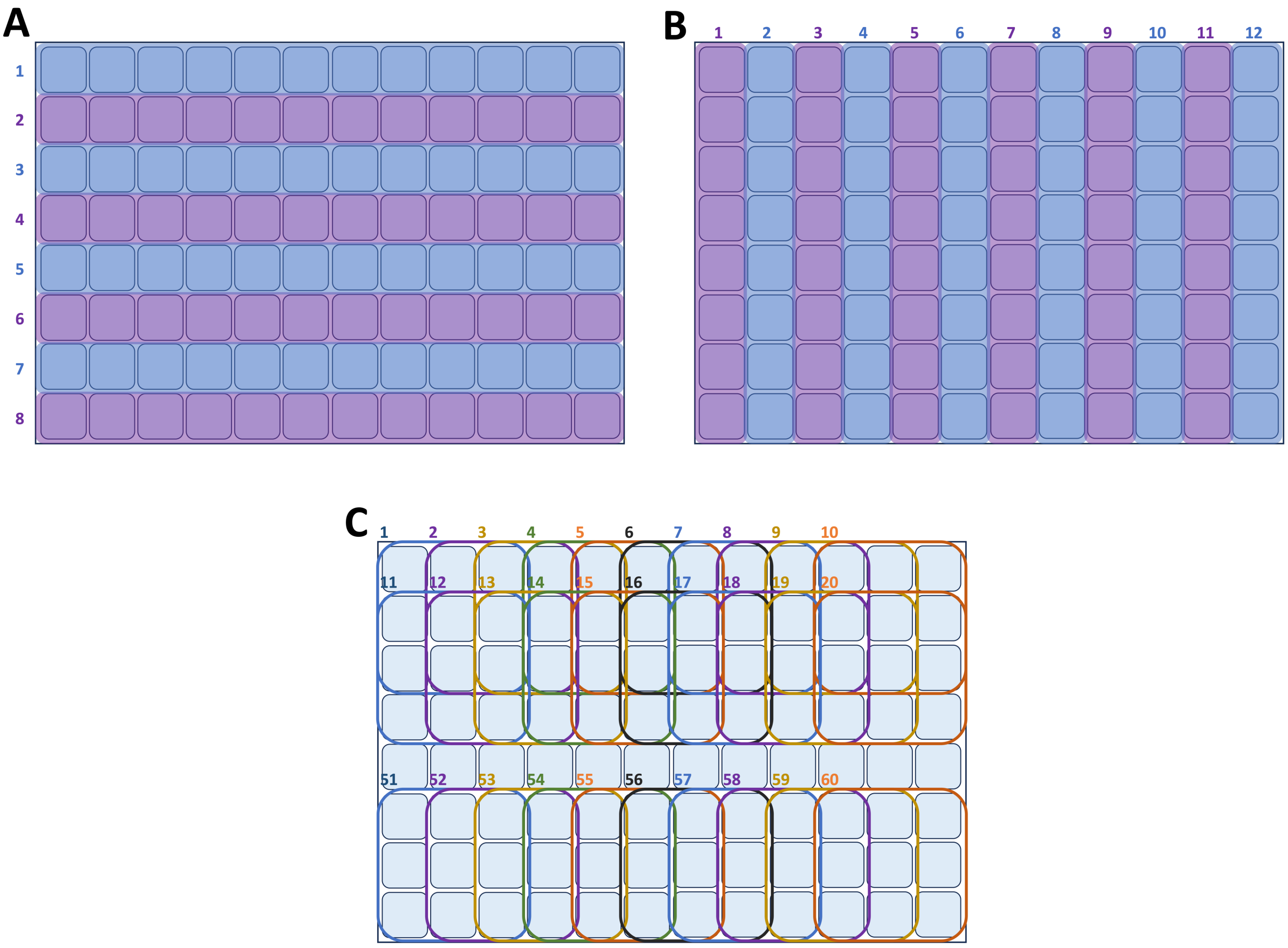

**Supplementary Fig. 2.** In a 96-well plate, we ignore wells set aside for technology-dependent internal controls and treat the plate as a matrix with approximately: **(A)** eight row indices ($rows = \{1 \dots 8\}$); **(B)** twelve column indices ($columns = \{1 \dots 12\}$); and **(C)** sixty patch indices ($patches = \{1 \dots 60\}$). Now, ignoring variable weightings, the maximal penalty for homogeneous rows and columns combined is $|rows| + |columns| \approx 8 + 12 \approx 20$, with an average maximal penalty of approximately 10. However, the corresponding maximal penalty for patches is $|patches| \approx 60$. Therefore, if we want the penalty due to patches to have the same importance as that due to either rows or columns, we must down-weight the patches penalty by $w_{pat} = \frac{1}{6}$ (because by $\frac{1}{6}.60 = 10$).

**Supplementary Figure 3. Pairwise well distances on a 96-well plate**

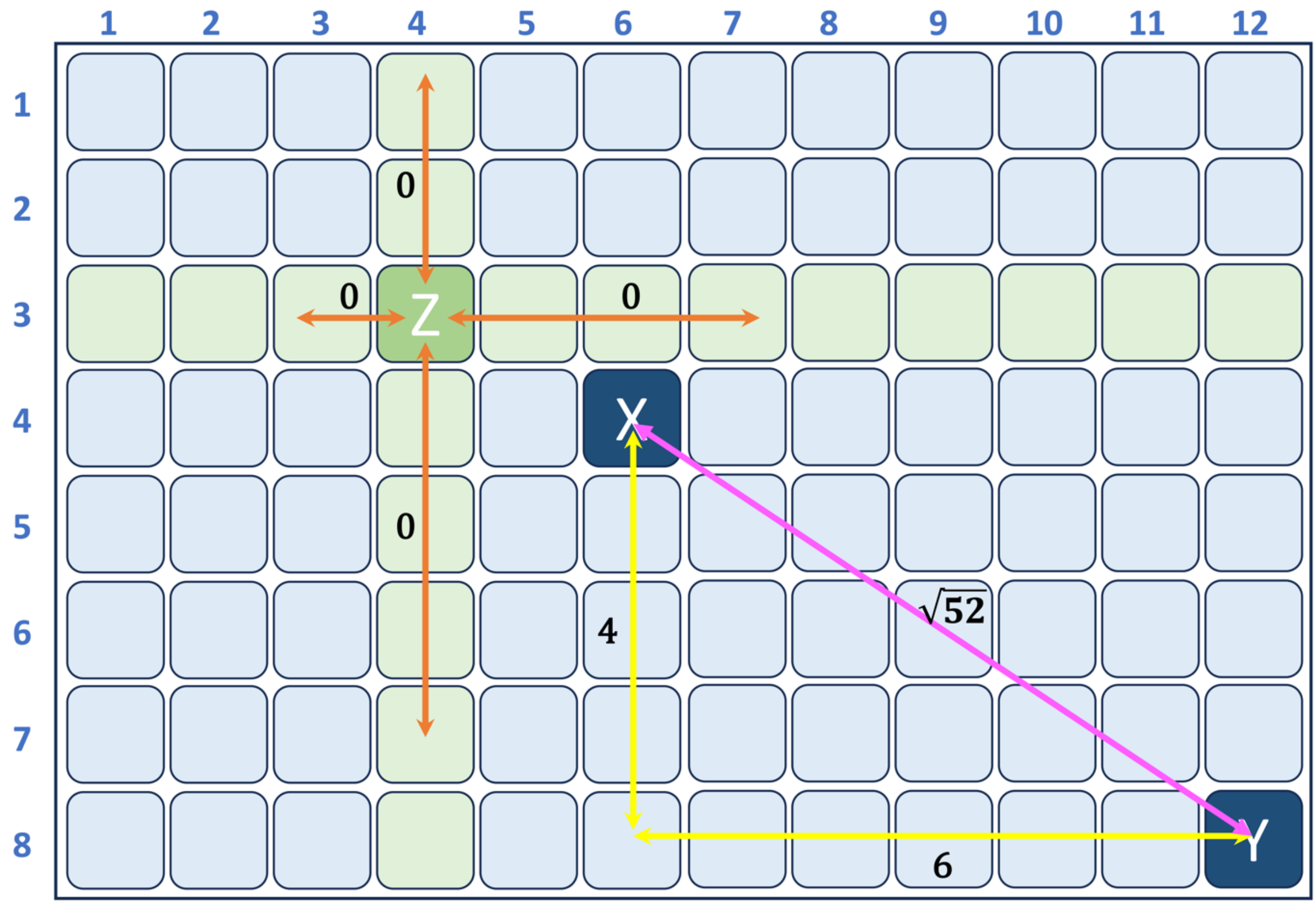

**Supplementary Fig. 3.** Wells sharing a row or column have a distance of zero, so well ‘Z’ (shaded in green) has a distance of zero to all the wells with which it shares a row or column (shaded in light green). In all other cases, the distance between two wells is calculated as the Euclidean distance between their row and column indices; here, wells X and Y have a distance of $\sqrt{4^2 + 6^2} = \sqrt{16 + 36} = \sqrt{52} \approx 7.2$.

**Supplementary Figure 4. Synthesised multivariate data for easyplater test**

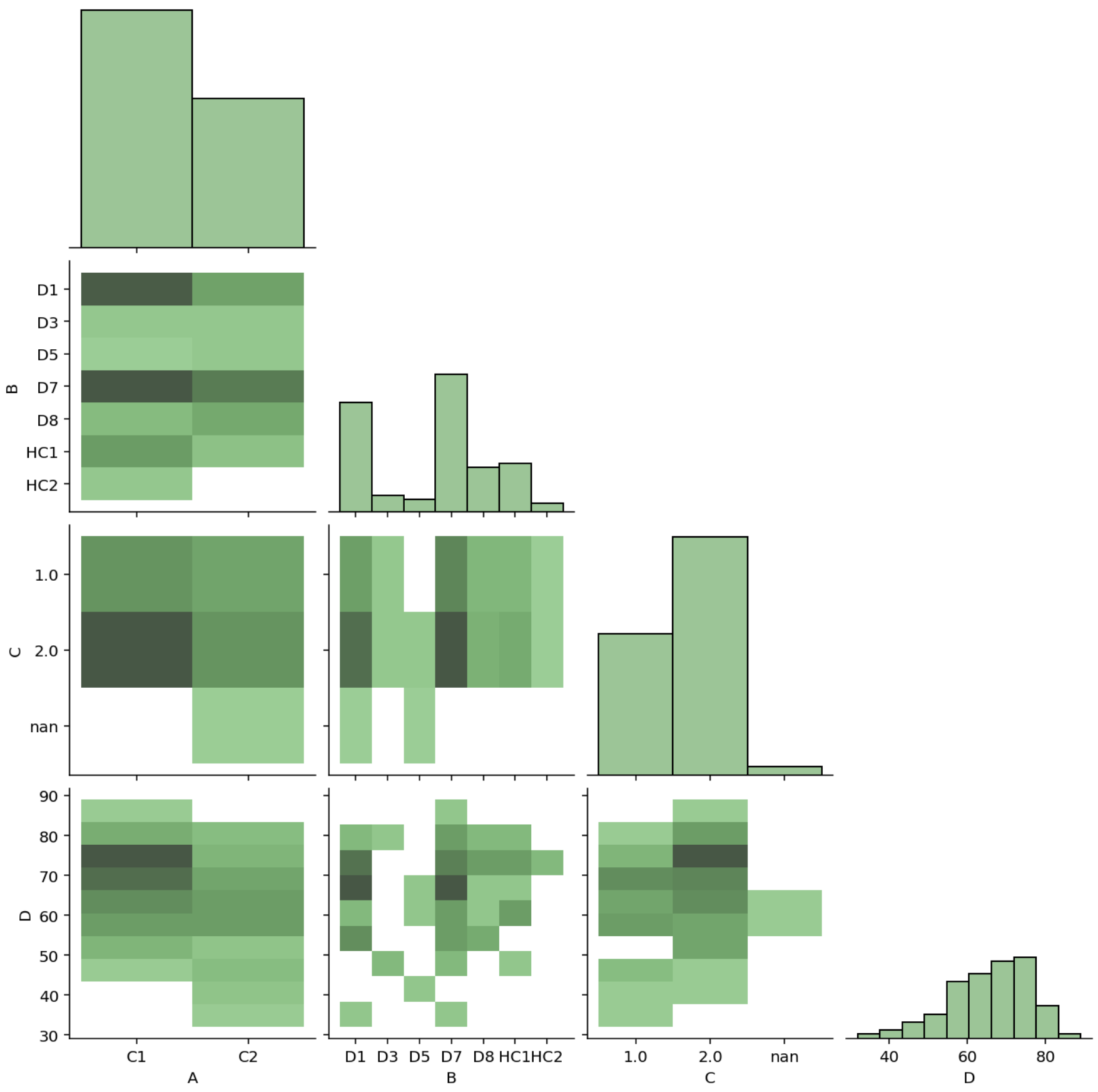

**Supplementary Fig. 4.** We synthesized multivariate data for 96 samples. Four variables were synthesized for each sample: variables A, B, and C were directly synthesized as categorical, whereas variable D was first synthesized as a continuous, integer variable and then subsequently binned into 10 groups prior to starting the main plate design process. Distributions for A, B, C, and D are shown on the diagonal. Pairwise joint distributions are shown in the lower half of the plot, with darker shades indicating a higher joint count. To synthesize this data, we used a subset of clinical data randomly drawn from an in-house dataset comprising >1000 samples each with >10 recorded clinical variables. To maintain anonymity of subjects and samples, all variable labels and categories have been changed and each clinical variable has been independently randomized amongst samples such that no sample level correlation structure persists in the data. In the final set of synthesized data, variables A and B have no missing data, and variables C and D have 2/96 and 5/96 missing values, respectively.

**Supplementary Figure 5. Additional distributions of plate design scores obtained for plate designs generated using random number generation, OmixeR, and easyplater**

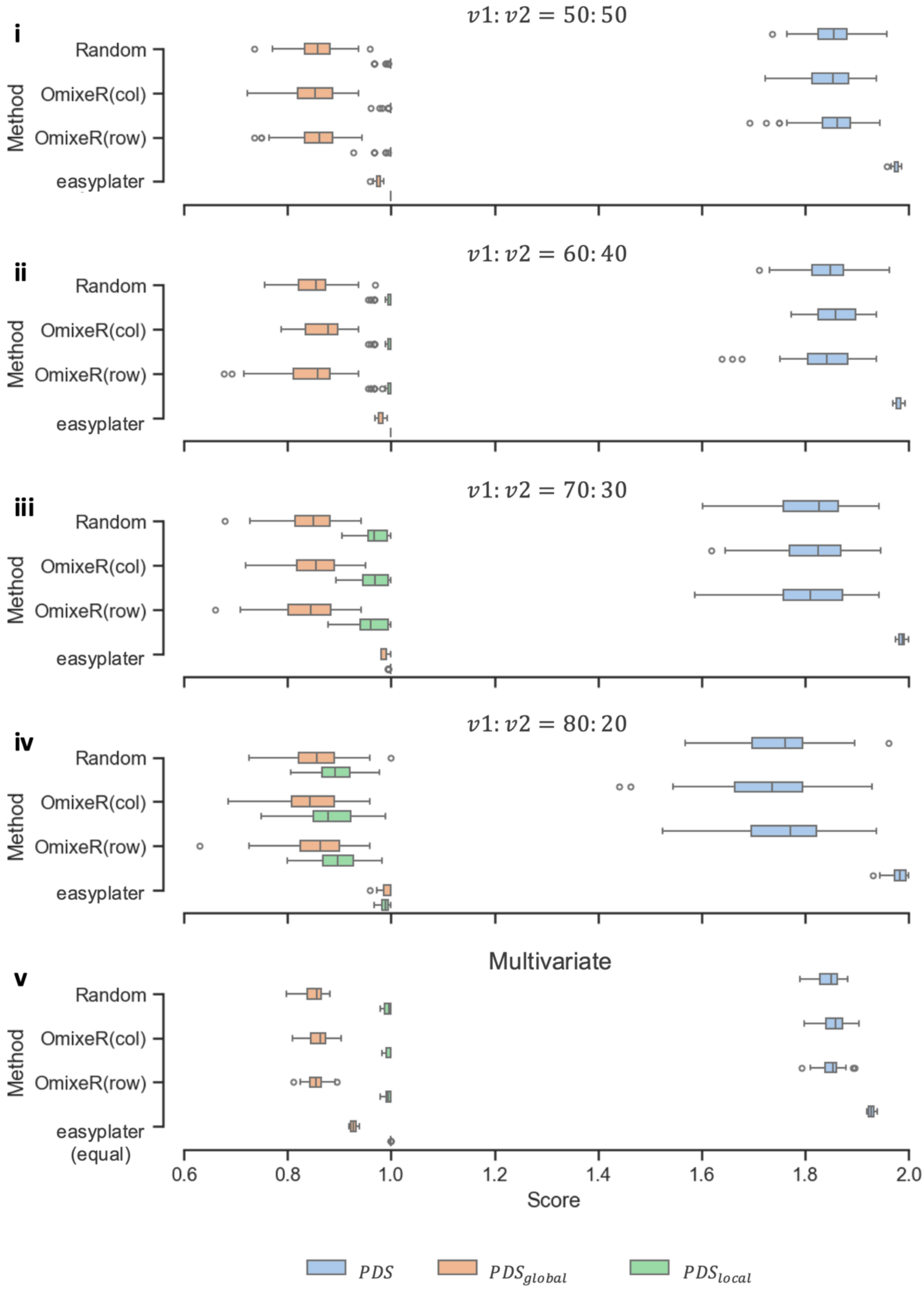

**Supplementary Fig. 5. (i)** Using random number generation, OmixeR, and easyplater, 50 plate designs were generated for a synthetic, 96-sample univariate dataset with a 50:50 *v1*:*v2* sample ratio split over variable *V*; Random = random number generator, OmixeR(col) = OmixeR *column* mode, OmixeR(row) = OmixeR *row* mode; blue, orange, and green boxplots show distributions of *PDS*, $PDS_{global}$, and $PDS_{local}$, respectively. **(ii, iii, and iv)** As for (i), but with a 60:40, 70:30, and 80:20 *v1*:*v2* ratio, respectively. **(v)** As for (i-iv), but each sample had four variables, *A*, *B*, *C*, and *D* (Supplementary Fig. S4), with equal weights assigned to variables in easyplater (option not available in remaining methods).

**Supplementary Figure 6. Contributions to improved *PDS* of Steps 2 and 3 of the easyplater plate design algorithm**

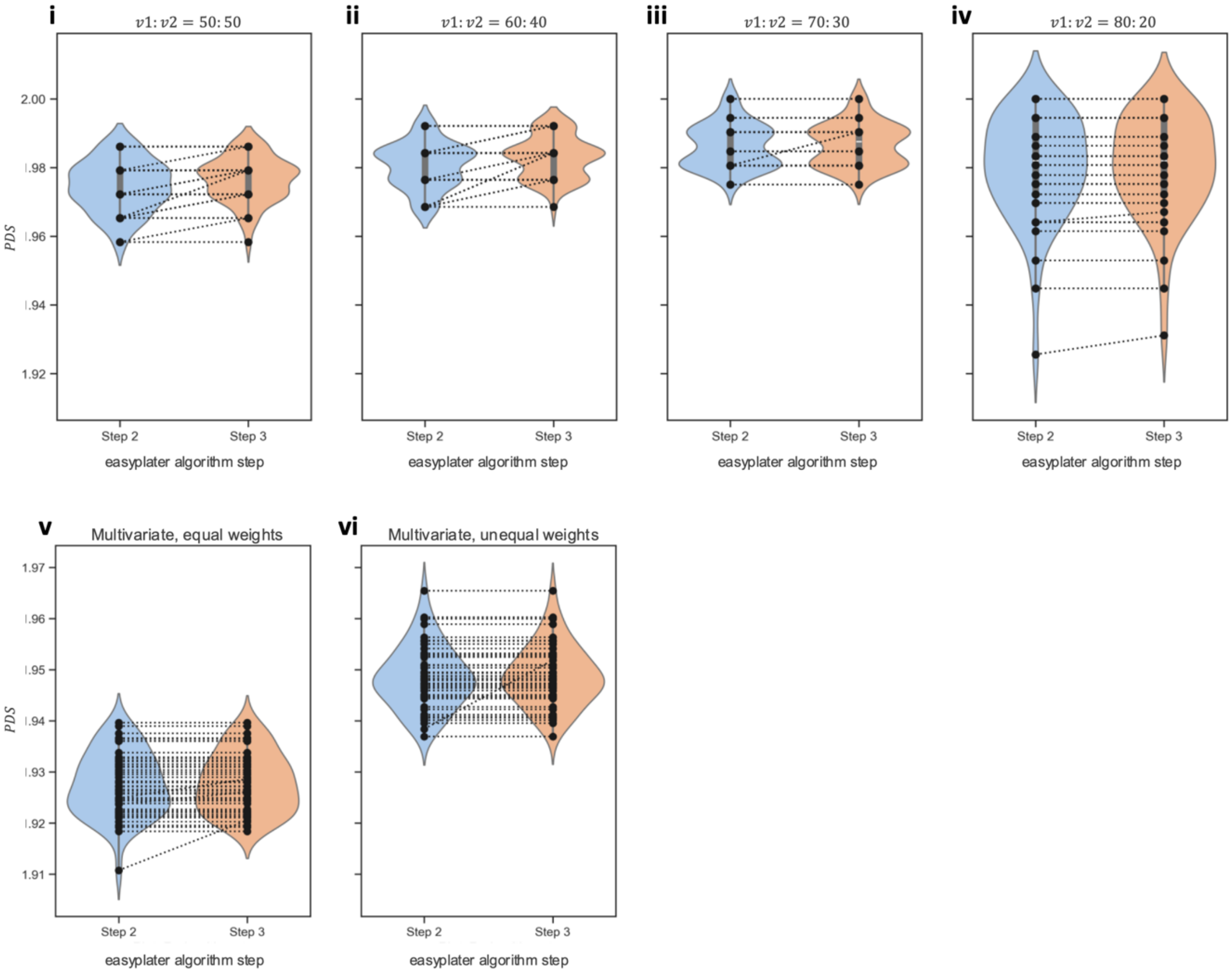

**Supplementary Fig. 6. (i)** Using easyplater, 50 plate designs were generated for a synthetic, 96-sample univariate dataset with a 50:50 *v1*:*v2* sample ratio split over variable *V*; blue and orange violin plots show the distribution of *PDS* after Step 2 and Step 3 of the plate design algorithm, respectively, with paired *PDS*s from Step 2 and 3 of each of the fifty plate designs shown as pairs of points connected by dotted lines overlayed on the violin plots. **(ii)**, **(iii)**, and **(iv)**: as for (i) but with 60:40, 70:30, and 80:20 *v1:v2* sample ratio splits, respectively. **(v)**: As for (i)-(iv), but plate designs were generated for a synthetic, 96-sample multivariate dataset where each sample had a recorded value for four variables, *A*, *B*, *C*, and *D* (Supplementary Fig. S5), with equal weights assigned to variables. **(vi)** As for (v) but with weights 0.1, 0.65, 0.15, and 0.1 assigned to *A*, *B*, *C*, and *D*, respectively. Note that results reported here are for the same designs that were generated for, and reported in, Fig. 2 and Supplementary Fig. S5. Note also that for any given easyplater layout, the *PDS* reported after Step 3 is the *PDS* for the final, outputted layout.

**Supplementary Figure 7. Comparing *PDS*, OmixeR(row), and Omixer(col) scores for 1000 random plate designs**

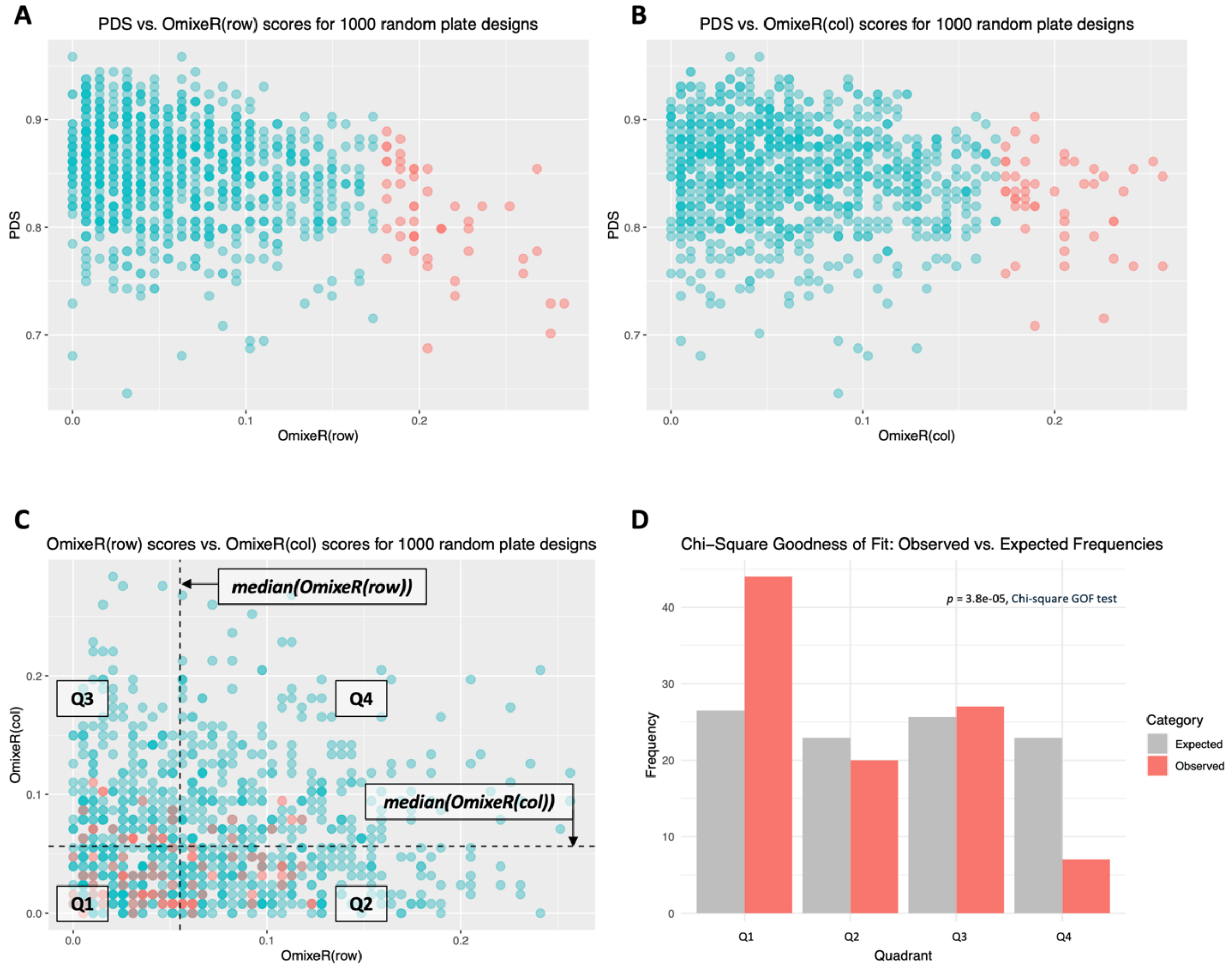

**Supplementary Fig. 7.** We generated 1000 plate designs using random number generation for a synthetic, 96-sample univariate dataset with a 50:50 *v1*:*v2* sample ratio split over variable *V*. For each of these random plate designs we calculated *PDS*, OmixeR(row), and OmixeR(col) scores. Since easyplater tries to maximize *PDS,* whilst OmixeR seeks to minimize the sum of absolute correlations between variables and plate zones, we would expect to see a strong negative correlation between these two measures if they were proxies of one another. **(A)** *PDS* vs. OmixeR(row) scores. OmixeR excludes plate designs with evidence for association between plate zones and outcome (in this case when $p < 0.05$ for association between rows and *v1*/*v2* status); here, points corresponding to excluded designs are red, otherwise green. **(B)**, as for (A), but *PDS* vs. OmixeR(col) scores. **(C)** OmixeR(col) vs. OmixeR(row) scores. Points are coloured red if they correspond to a randomized design with a 'high' *PDS*, here defined as *PDS* > 90$^{th}$ percentile of all 1000 *PDS*s obtained; otherwise, green. The plot is split into quadrants using the median OmixeR(row) score and the median OmixeR(col) score, drawn as vertical and horizontal dashed lines, respectively: Quadrant 1 (Q1) is defined as all designs scoring OmixeR(row) <= median(OmixeR(row)) AND OmixeR(col) <= median(OmixeR(col)); Quadrant 2 (Q2) is defined as all designs scoring OmixeR(row) > median(OmixeR(row)) AND OmixeR(col) <= median(OmixeR(col)); Quadrant 3 (Q3) is defined as all designs scoring OmixeR(row) <= median(OmixeR(row)) AND OmixeR(col) > median(OmixeR(col)); and Quadrant 4 (Q4) is defined as all designs scoring OmixeR(row) > median(OmixeR(row)) AND OmixeR(col) > median(OmixeR(col)). **(D)** Bar chart of the number of observed (red) vs. expected (grey) high *PDS* designs in each of quadrants Q1-Q4 of scatter plot (C). A Chi-square Goodness of Fit test (GoF test) yields a *p-value* of 3.8e-05, indicating that we should reject the null hypothesis that high *PDS* designs are present in the same proportion across Q1-Q4, in favour of the alternative hypothesis that they are not. We checked to see how sensitive this result was to changes in thresholds for Q1-Q4, as well as changing the threshold for a 'high' *PDS* score. In particular, we defined quadrants of (C) using the 40$^{th}$ percentile of OmixeR(row) and OmixeR(col) scores, and ran this analysis three times with high *PDS* defined as: > 70$^{th}$ percentile of *PDS*s; > 80$^{th}$ percentile of *PDS*s; and > 90$^{th}$ percentile of *PDS*s. We then repeated the analysis using the same three thresholds for high *PDS*, but defining Q1-Q4 using the 60$^{th}$ percentile of OmixeR(row) and OmixeR(col) scores. Finally, we re-ran our original analysis with Q1-Q4 defined using median OmixeR(row) and OmixeR(col) scores, but with high *PDS* defined as >70$^{th}$ percentile of *PDS*s and >80$^{th}$ percentile of *PDS*s. All nine GoF tests yielded a *p-value* < 5.0e-04, indicating that this result is robust and not sensitive to changes in thresholds.

## Supplementary Boxes

**Supplementary Box 1. A toy example to show how we calculate $PDS_{global_{v_x}}$, the sub-score of $PDS_{global_v}$ accounting for the randomization of value *x* of *v* across a plate design**

Consider the toy example of designing an 18-well microplate layout for 18 samples, where each sample has one recorded variable, *shade*. Variable *shade* has two categorical values, *light* and *dark*; half the samples are *light* and the other half are *dark*. Here, we show how to calculate $PDS_{global_{shade_dark}}$, the sub-score of $PDS_{global_{shade}}$ accounting for the spatial configuration of value *dark* across the plate.

**(1)** Start by counting $t_{shade_dark}$, the number of *dark* samples assigned to row and column exclusive wells. Do this by constructing a network in which the nodes are wells containing *dark* samples, and pairs of nodes are linked by an edge when the corresponding pair of wells are row and column exclusive of one another. Represent this network as a matrix and count the number of 1s in the lower (or upper) triangle of the matrix.

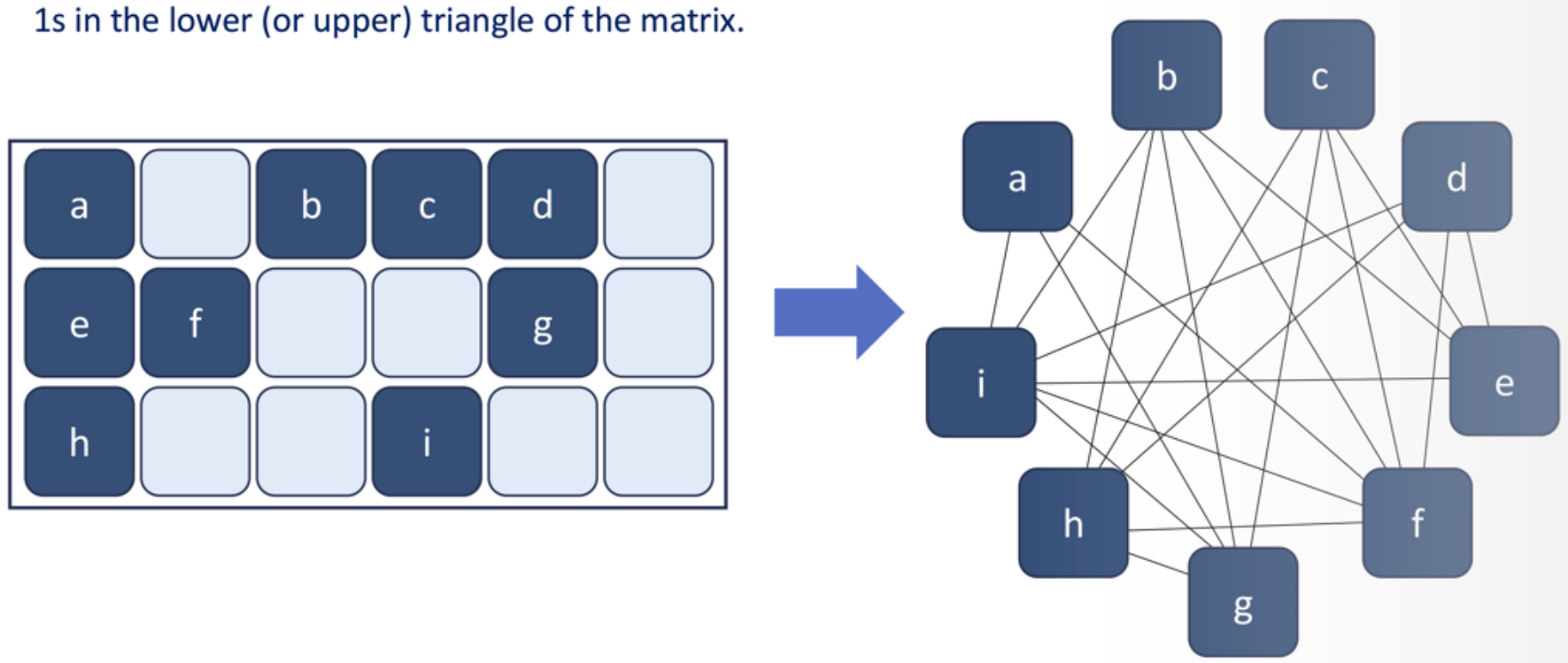

| | a | b | c | d | e | f | g | h | i |
|---|---|---|---|---|---|---|---|---|---|
| a | 0 | 0 | 0 | 0 | 0 | 1 | 1 | 0 | 1 |
| b | 0 | 0 | 0 | 0 | 1 | 1 | 1 | 1 | 1 |
| c | 0 | 0 | 0 | 0 | 1 | 1 | 1 | 1 | 0 |
| d | 0 | 0 | 0 | 0 | 1 | 1 | 0 | 1 | 1 |
| e | 0 | 1 | 1 | 1 | 0 | 0 | 0 | 0 | 1 |
| f | 1 | 1 | 1 | 1 | 0 | 0 | 0 | 1 | 1 |
| g | 1 | 1 | 1 | 0 | 0 | 0 | 0 | 1 | 1 |
| h | 0 | 1 | 1 | 1 | 0 | 1 | 1 | 0 | 0 |
| i | 1 | 1 | 0 | 1 | 1 | 1 | 1 | 0 | 0 |

$$t_{shade_dark} = 21$$

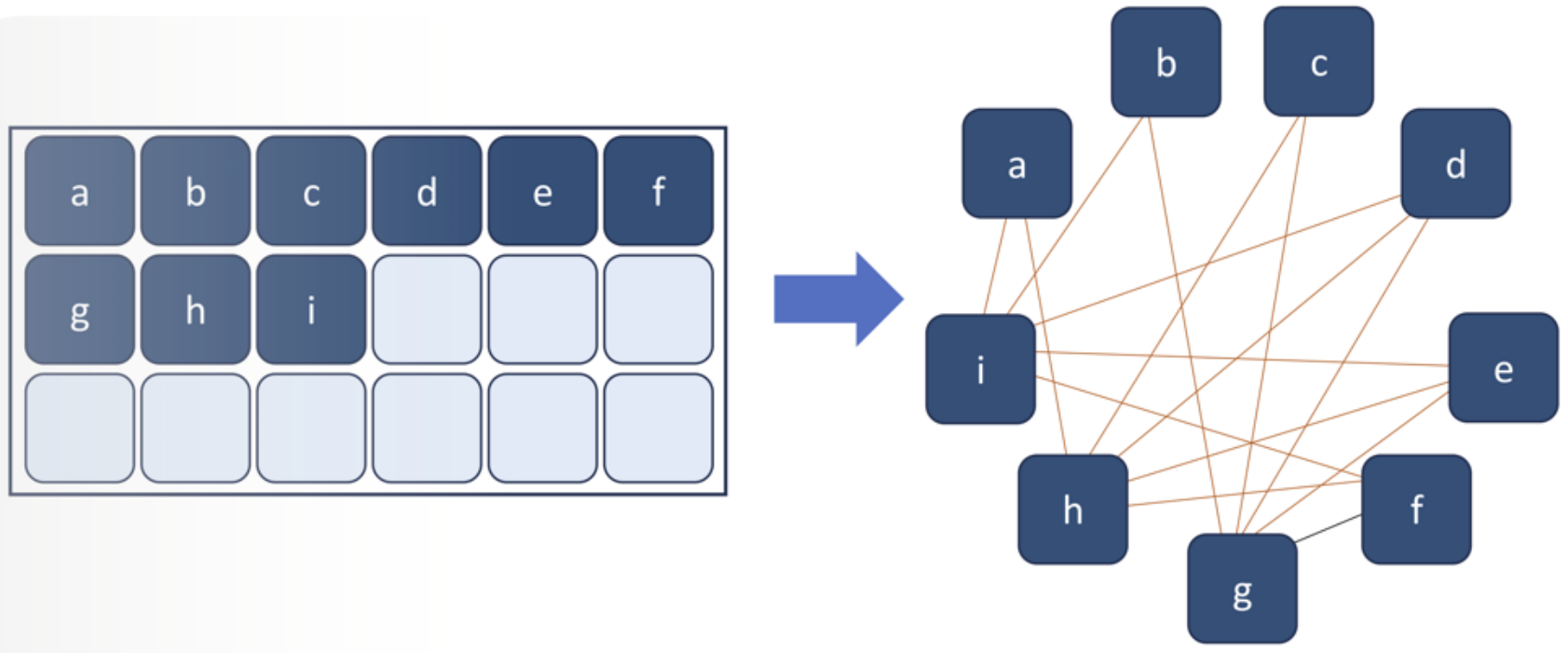

$$t_{shade_dark_{min}} =$$

$$edgesBetweenSamplesInFullRows + edgesTo/FromSamplesInPartialRow =$$

a b c d e f: $\frac{(fullRows.cols)((fullRows-1).(cols-1))}{2}$ → $\frac{(1*6)((1-1)(6-1))}{2}$

g h i: $samplesInPartialRow.fullRows(cols-1)$ → $3*1(6-1)$

$$0 + 15 = 15$$

**(2)** Calculate $t_{shade_dark_{min}}$, the minimum number of pairs of *dark* samples that could feasibly be assigned to row and column exclusive wells in this 3 by 6 18-well plate. To do this, we use integer division and the modulo function to find the minimum number of full rows that *dark* samples could occupy, and hence the number of *dark* samples in full rows; we also find the number of *dark* samples in the last, partially filled row. Next, we **imagine*** a network in which the nodes are wells containing *dark* samples, and pairs of nodes are linked by an edge when the corresponding pair of wells are row and column exclusive of one another. We then use this idea to build an equation for the number of edges between *dark* samples in full rows; similarly, we build an equation for the number of edges linking *dark* samples in the partial rows to/from *dark* samples in the full rows. $t_{shade_dark_{min}}$ is then the sum of these subtotals.

**We don't actually need to construct this network or store it in memory.*

**Supplementary Box 1. (continued)**

**(3)** Calculate $t_{shade_dark_{max}}$, the maximum number of pairs of *dark* samples that could feasibly be assigned to row and column exclusive wells in this 3 by 6 18-well plate. To do this, we start by imagining that we actually have an infinite sized microplate, such that every *dark* sample can be placed in a well that is row and column exclusive of every other well containing a *dark* sample. Then, our network, in which nodes are wells containing *dark* samples and edges link row and column exclusive wells, is fully connected. Now, we know that the number of edges in a fully connected network with $n$ nodes is $\frac{n(n-1)}{2}$, so we calculate $\frac{t_{shade_dark}(t_{shade_dark}-1)}{2}$ and then find $t_{shade_dark_{max}}$ by subtracting the number of pairs of *dark* samples in row-sharing and column-sharing wells (calculated using integer division and the modulo function).

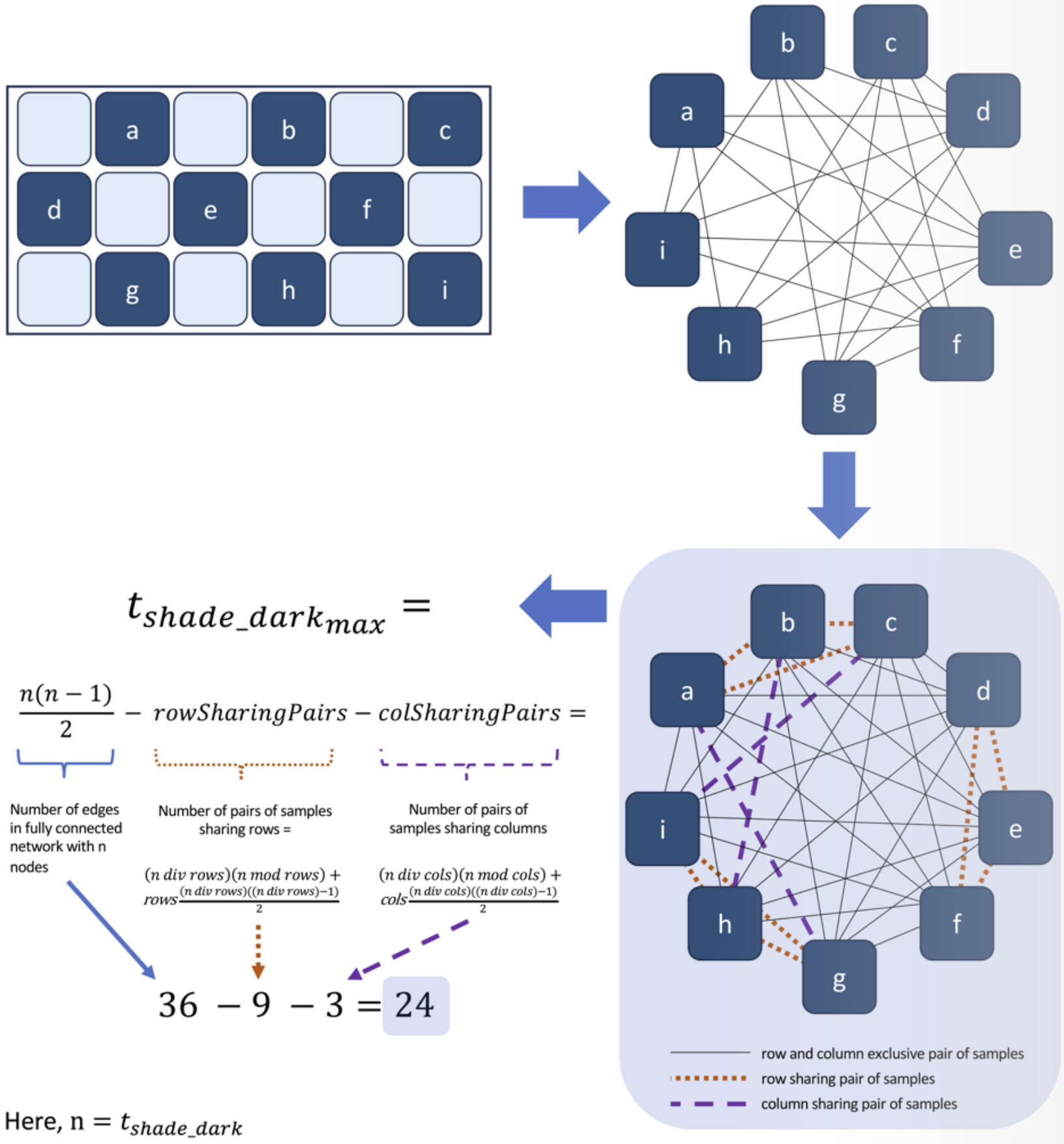

**(4)** Finally, calculate $PDS_{global_{shade_dark}}$ by shifting and scaling $t_{shade_dark}$ to range [0,1] using $t_{shade_dark_{min}}$ and $(t_{shade_dark_{max}} - t_{shade_dark_{min}})$ as the shift and scale constants, respectively.

$$PDS_{global_{shade_dark}} = \frac{(t_{shade_dark} - t_{shade_dark_{min}})}{(t_{shade_dark_{max}} - t_{shade_dark_{min}})}$$

$$= \frac{(21-15)}{(24-15)} = \frac{6}{9}$$

$$= 0.66$$

**Supplementary Box 2. Identifying well-sets wherein all pairwise well distances exceed some threshold**

**(1)** Consider a toy example of a microplate with just 24 wells:

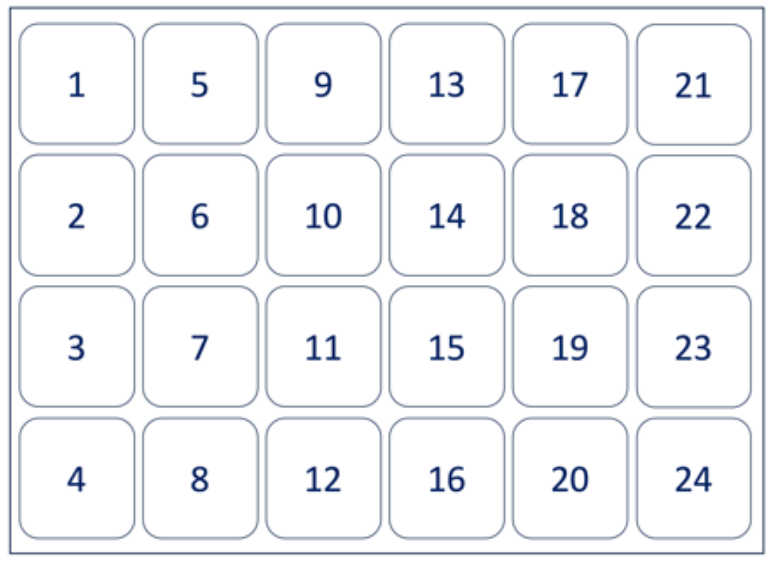

**(2)** If we look at each well in turn (dark blue), then we can identify:

(i) all wells a distance of 0 away from the well (green),
(ii) all wells a distance of $\sqrt{2}$ away from the well (yellow),
(iii) all wells a distance of $\sqrt{5}$ away from the well (peach),
(iv) all wells $\geq \sqrt{10}$ away from the well (light blue).

In this toy example, we have set the 'distal' threshold at $\geq \sqrt{10}$, meaning that for any given dark blue well, its distal wells are light blue wells.

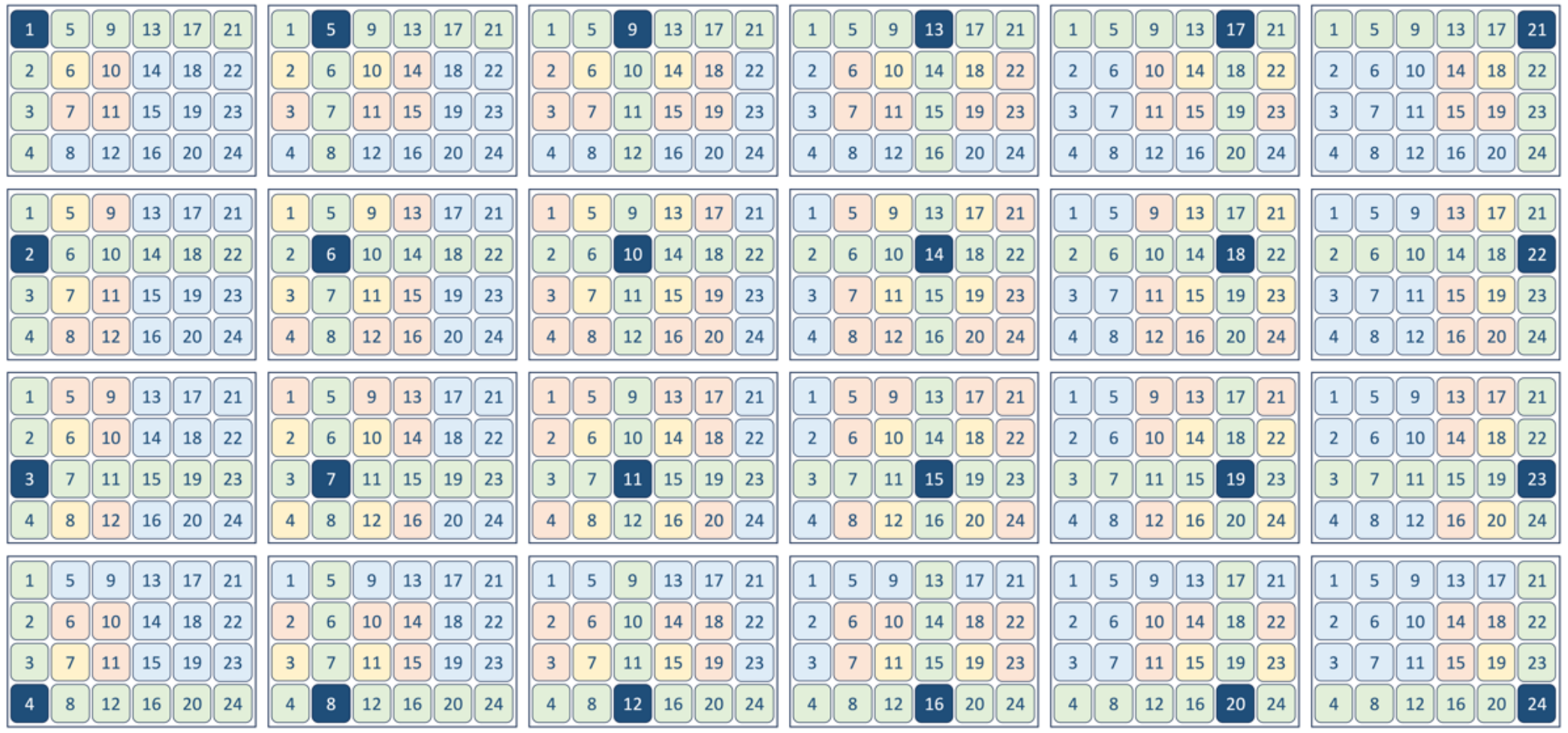

**(3)** Build a network in which nodes represent wells, and pairs of nodes are linked by an edge when the corresponding pair of wells are distal to one another.

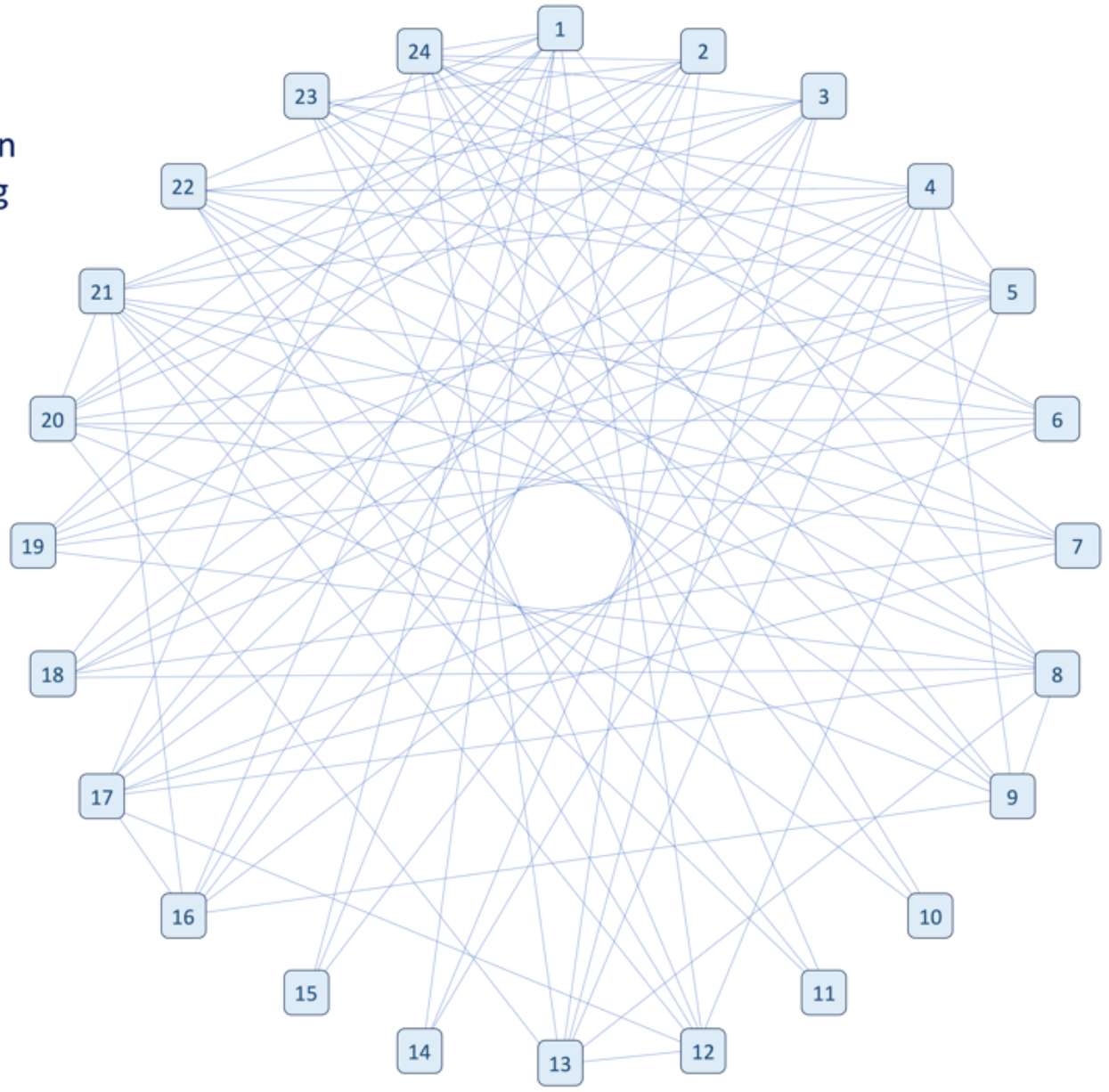

**(4)** Find cliques of nodes in the network; these cliques are then well-sets wherein all pairwise well distances exceed some threshold, in this case $\geq \sqrt{10}$ .

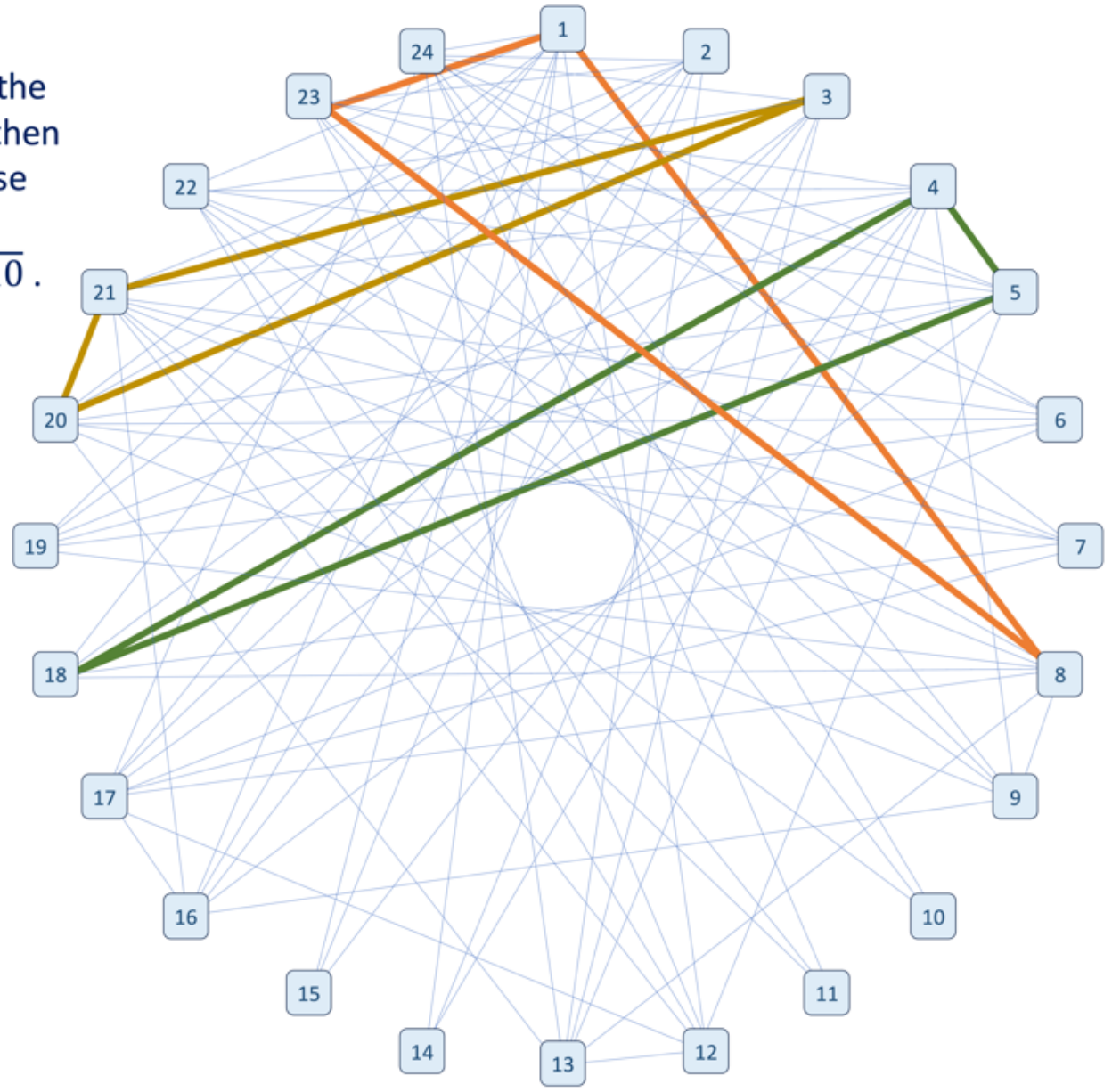

**Supplementary Box 3. Generating a set of mutually exclusive sample switches and performing simultaneous multiple-sample switching**

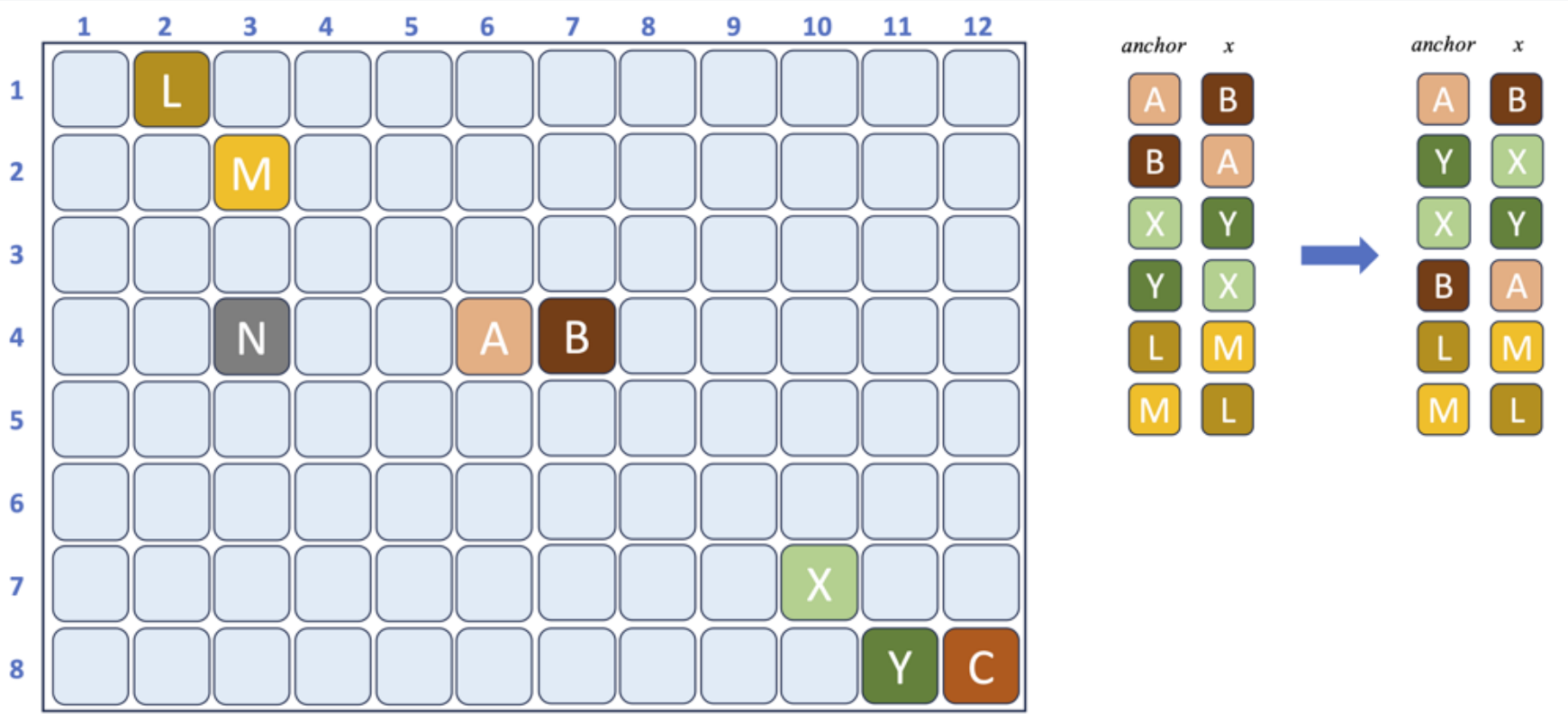

**(1) Build a randomly ordered list of *splitting* pairs: pairs of 'similar' samples in 'nearby' wells.** Go through a plate design and find all pairs of samples which are 'similar' and in 'nearby' wells; each pair is listed twice such that member wells take either the *anchor* label or the *x* label. Finish this step by randomizing the order of the list of *splitting* pairs.

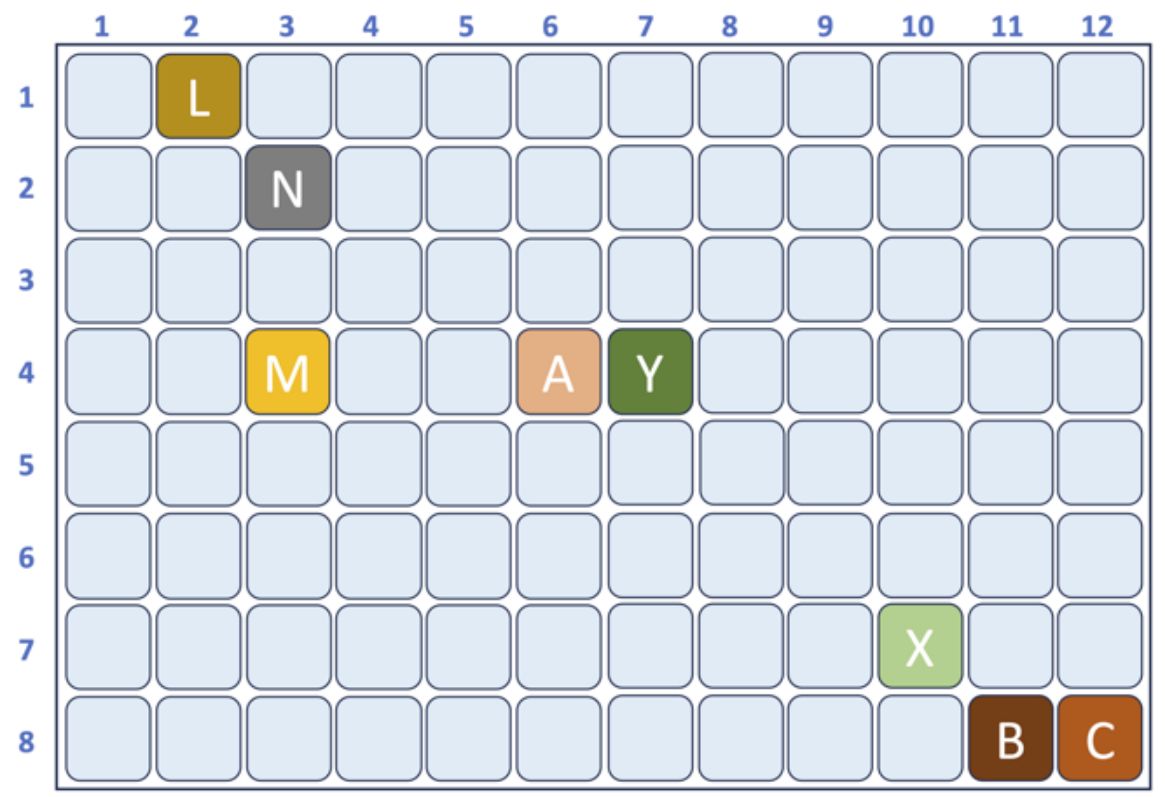

**(3) The new plate design, after samples B & Y, and M & N are simultaneously switched, in one step.** Note that when using the easyplater package, users can set thresholds for 'similarity', 'dissimilarity', 'nearby', and 'distal'; default values are similarity>0.5, similarity<=0.5, distance<1, and distance>6, respectively. For ease of depiction, here, thresholds for 'nearby' and 'distal' are distance<1.5 and distance>2, respectively.

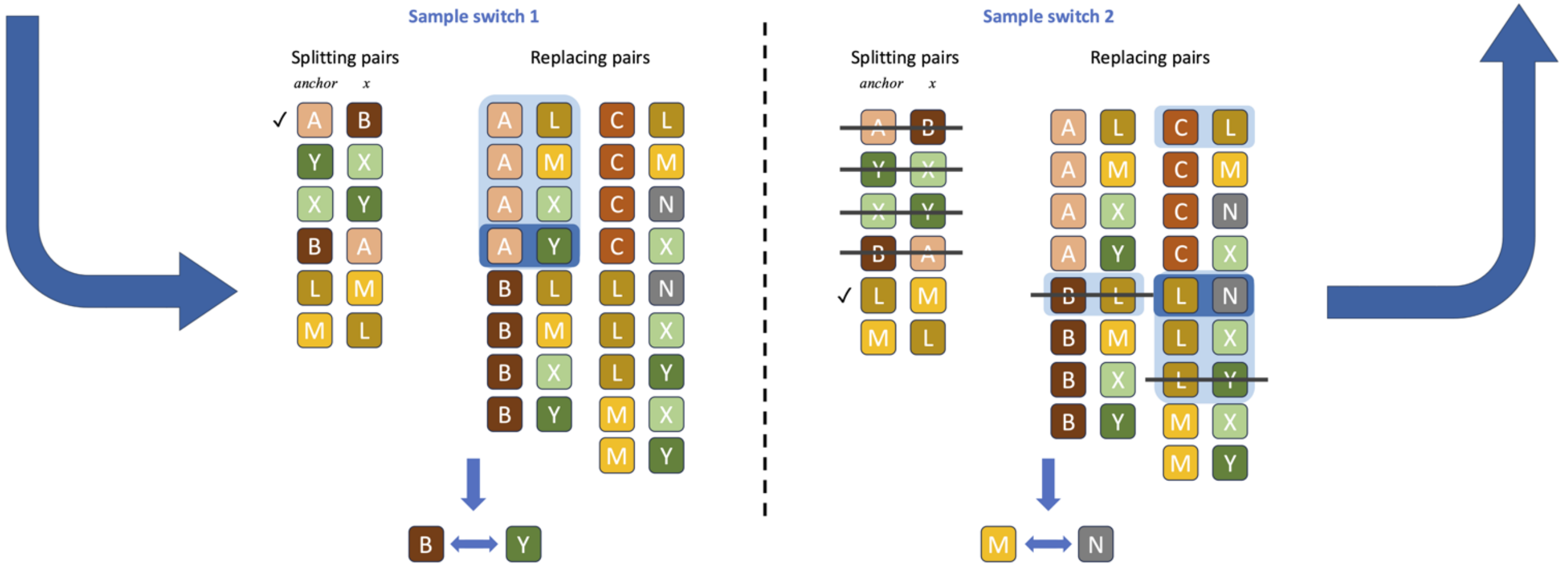

**(2) For each *splitting* pair, in turn, identify potential *replacing* pairs:** pairs of 'dissimilar' samples in 'distal' wells such that one of the samples matches the *anchor* sample from the *splitting* pair, and such that the remaining sample *y* has not been selected for an earlier switch. Here, in Sample switch 1, *splitting* pair (*anchor*=A, *x*=B) has been chosen, and its list of potential *replacing* pairs (amongst all pairs of dissimilar samples in distal wells) are highlighted with a light blue background (amongst all pairs of dissimilar samples in distal wells); here (*anchor*=A, *y*=Y) has been randomly picked (dark blue background). Thus, samples *B* and *Y* have been identified for switching. Next, in Sample switch 2, *splitting* pair (*anchor*=L, *x*=M) has been chosen because all *splitting* pairs listed before it contain a sample already identified for switching (i.e., either B or. Y). Again, the list of possible *replacing* pairs are highlighted (light blue), but this time *replacing* pairs where one sample has been selected for an earlier switch are crossed out. Finally, *replacing* pair (*anchor*=L, *y*=N) has been randomly picked (dark blue); thus samples M and N have been identified for switching.

## Supplementary Tables

**Supplementary Table 1. Mean runtimes for plate design algorithms on a bioinformatics-capable laptop**

| Data | Random | OmixeR(col) | Omixer(row) | easyplater |
|---|---|---|---|---|
| 50:50 | 4.9e-05s | 33.6s | 29.9s | 18.6s |
| 60:40 | 5.3e-05s | 35.8s | 31.6s | 16.4s |
| 70:30 | 5.1e-05s | 33.1s | 29.6s | 15.3s |
| 80:20 | 4.8e-05s | 33.0s | 29.4s | 16.8s |
| Multi: Even | 5.1e-05s | 41.6s | 36.0s | 36.1s |
| Multi: Uneven | 4.3e-05s | 35.7s | 35.0s | 104.5s |

**Supplementary Table 1.** Using random number generation, OmixeR, and easyplater, 50 plate designs were generated for a synthetic, 96-sample univariate dataset with a 50:50 *v1:v2* sample ratio split over variable *V*. We repeated this process for univariate data with 60:40, 70:30, and 80:20 *v1:v2* sample ratio splits over variable *V*. Finally, we repeated this process again, but this time with a synthetic, 96-sample multivariate dataset where each sample had a recorded value for four variables, *A*, *B*, *C*, and *D* (Supplementary Fig. S5); once with even weights assigned to variables in easyplater, and once with uneven weights (0.1, 0.65, 0.15, and 0.1) assigned to variables in easyplater (the other algorithms do not provide functionality to differentially weight variables). The results from these analyses are discussed in the main text (Section 3: Results), and shown in Fig. 2, Supplementary Fig. S6, and Supplementary Fig. S7. Here, we give the mean runtimes, over 50 runs, for generating plate designs for each data set with each algorithm. Runtimes were measured on a MacBook Pro laptop with an Apple M3 Pro CPU and 36GB RAM. **Abbreviations:** 50:50, 60:40, 70:30, and 80:20 = mean runtimes for designing plates for the 96-sample univariate datasets with 50:50, 60:40, 70:30, and 80:20 *v1:v2* sample ratio splits, respectively. Multi: Even, and Multi: Uneven = mean runtimes for designing plates for the 96-sample multivariate data with even and uneven weights, respectively. Random = random number generation, OmixeR(col) = OmixeR *column* mode, OmixeR(row) = OmixeR *row* mode.